\documentclass[12pt]{article} 

\usepackage{physics}
\usepackage{graphicx}
\usepackage{amsmath}
\usepackage{amssymb}
\usepackage{hyperref}
\usepackage[paperwidth=8.5in,paperheight=11in,textwidth=6.25in,textheight=9.1in,
 centering]{geometry}

\usepackage{cite}
\usepackage{amscd}
\usepackage{setspace}
\usepackage{bbm}
\usepackage{dsfont}
\usepackage{cancel}
\usepackage{relsize}
\usepackage[utf8]{inputenc}
\usepackage{mathtools}
\usepackage{newtxtext,newtxmath}
\numberwithin{equation}{section}

\begin{document}
\begin{titlepage}

\begin{flushright}

\end{flushright}

\vskip 3cm

\begin{center}
{\Large \bf
On mixed 't Hooft anomalies of emergent symmetries}
\vskip 2.0cm

 Wei Gu$^1$, Du Pei$^2$, and Xingyang Yu$^3$  \\

\bigskip
\begin{tabular}{cc}
 $^1$ Zhejiang Institute of Modern Physics, School of Physics,  Zhejiang University\\
Hangzhou, Zhejiang 310058, China\\

$^2$ Centre for Quantum Mathematics, University of Southern Denmark \\
Campusvej 55, Odense 5230, Denmark\\

$^3$ Department of Physics, Robeson Hall,
Virginia Tech\\
Blacksburg, VA 24061, USA
 \end{tabular}

\vskip 1cm

\textbf{Abstract}
\end{center}

In this paper, we investigate the dynamical constraints imposed on the UV theory when it develops an emergent symmetry in the infrared with mixed 't Hooft anomalies. We demonstrate that, under certain conditions, the UV theory must contain non-genuine operators. Our primary examples illustrating this phenomenon are 2D gauged linear sigma models and 3D Chern-Simons-matter theories. Through this analysis, we establish connections between different classes of topological quantum field theories and propose a correspondence between quantum cohomologies of distinct target spaces.

\medskip
\noindent

\bigskip
\vfill
\end{titlepage}

\setcounter{tocdepth}{2}
\tableofcontents
\section{The motivation}\label{Intr}

The 't Hooft anomaly was initially introduced by 't Hooft to contrain the low-energy dynamics of a UV theory \cite{tHooft:1979rat}, which recently has been widely applied to systems with generalized symmetries\cite{Gaiotto:2014kfa}, leading to the discovery of new structures. In all of the application scenarios, the anomaly is almost always used to probe infrared physics. However, in this paper, we are interested in asking about the opposite direction:
\begin{center}
   \textit{Can we probe the UV dynamics from given IR data utilizing the 't Hooft anomaly? } 
\end{center}

When the symmetry is emergent in the IR, this question might initially sound vacuous -- after all, numerous vastly different UV theories could flow to the same IR theory. Surprisingly, however, meaningful information about the UV theory can still be extracted under certain conditions. 

In this paper, we will restrict our attention to the following setting: 1) the infrared is a gapped phase, 2) it is described by a TQFT with non-trivial symmetries,\footnote{Here, by `TQFT,' we mean a theory whose operators and correlation functions are purely topological. They may not satisfy the strict Atiyah–Segal axioms of a TQFT. In particular, we do not require the theory to be defined on an arbitrary curved manifold, independent of the metric. For example, this would allow non-topological invertible theories, such as gravitational Chern--Simons theory in 3D, which commonly arise when integrating out fermions.} and 3) both the generators and charged operators associated with IR symmetries originate from the UV theory.\footnote{This assumption excludes scenarios where the IR theory already constitutes a complete UV description, where the anomalous IR symmetry is genuinely emergent and lacks a UV precursor, or where the anomaly is accounted for by the addition of decoupled UV/topological sectors rather than by the intrinsic UV dynamics. } Furthermore, when the UV theory has global symmetries, these can provide more specific labeling for the IR topological symmetry operators and/or their associated charged operators, allowing for better control over the operator spectrum under RG flow.

These conditions are satisfied by a wide range of physical systems -- including gauge theories, sigma models, and Chern-Simons-matter theories -- and enable precise analysis of the symmetry behavior under RG flow. In particular, we will show that they imply that the mixed 't Hooft anomaly in the IR can sometimes constrain the properties of UV operators.

Our strategy involves studying a series of concrete examples in which the IR theory is known, the anomaly structure is explicit, and the UV completion is under control. These include the following closely related theories: a) The compact BF theory and its UV completion via the gauged linear sigma models (GLSMs), as well as the non-linear sigma model (NLSMs) on projective or weighted projective spaces. b) The 2D $G/G$ WZW theory, its Verlinde algebra, and the corresponding quantum cohomology. c) The 3D Chern-Simons-matter theory and its associated quantum K-theory.

Through these examples, we observe a recurring phenomenon: an IR mixed anomaly involving an emergent symmetry necessarily requires the existence of certain non-genuine operators in the UV. These operators -- though the IR symmetry itself is not manifest in UV -- are essential in realizing the emergent symmetry in the IR. In several cases, these operators can be constructed explicitly using ingredients of the UV Lagrangian.

This leads us to a fundamental insight regarding topological field theories:

\begin{center}
    \textit{Witten-type TQFTs serve as crucial intermediaries  in the RG flow \\ from UV dynamics to IR topological phases}
\end{center}

Whereas the infrared phase often admits a ``Schwarz-type" TQFT description (satisfying axiomatic frameworks à la Atiyah--Segal, Lurie, etc.), Witten-type TQFTs—constructed via topological twist \cite{Witten:1988ze, Witten:1988xj}—provide the missing link. Their role is exemplified in Witten's seminal correspondence \cite{Witten:1993xi}:

\begin{center}
    Quantum cohomology of $Gr(k,N)$ $\cong$ Verlinde algebra of $U(k)_{N-k}$
\end{center}

Yet, this equivalence presents a symmetry mismatch puzzle:
\begin{itemize}
    \item Compact BF theory \cite{Blau:1993tv} , equiped with Verlinde algebra, has $\mathbb{Z}_N^{(0)} \times \mathbb{Z}_N^{(1)}$ symmetry
    \item NLSM on $\mathbb{P}^{N-1}$ has $SU(N)$ flavor symmetry + R-symmetries
\end{itemize}

The resolution lies in non-genuine UV operators that mediate symmetry structures under RG flow. As we demonstrate in Section \ref{GLSM}, the BF theory's 1-form symmetry emerges from GLSM composites $\Phi_i W[1]$, while its axial symmetry descends from UV axial R-symmetry. This establishes apparent symmetry differences that reflect complementary perspectives along the RG flow trajectory.

Beyond this specific case, we establish broader correspondences:
\begin{center}
    \text{Quantum cohomology of $\mathbb{CWP}(\vec{w})$} $\cong$ \text{Verlinde algebra of $U(1)_N$} \\
    \small{\text{with} $\overline{\Lambda}^N = \prod w_i$ and $N = \sum w_i$}
\end{center}

Remarkably, this equivalence depends only on $N = \sum w_i$ (Section \ref{CBQH}), revealing that global symmetries—not target space geometry—govern topological equivalences. The 3D Chern-Simons matter realization (Section \ref{CSMT}) further demonstrates how non-genuine operators mediate anomaly matching across dimensions.

Building on these examples and the fact that charged operators—unlike the symmetry generators—can be modified under RG flow, we propose, in a self-contained Section \ref{sec: general discussion},  two typical mechanisms for IR emergent symmetries inherited from UV:
\begin{itemize}
    \item \textbf{Mechanism I}: UV non-genuine composites $\leadsto$ IR symmetry-charged operators
    \item \textbf{Mechanism II}: UV non-genuine operators $\leadsto$ IR symmetry generators 
\end{itemize}
There exists a situation where the IR symmetry generators themselves are charged operators—a scenario that can only occur in odd dimensions. This general proposal resolves the puzzle of symmetry mismatches while demonstrating how IR anomalies constrain UV operator content. Our analysis culminates in a classification principle for TQFTs with equivalent mixed anomalies, and we anticipate numerous future applications of these results.

\emph{Note added:} While this work was being finalized and presented by one of the authors (WG) at \cite{WG:2025talk}, an independent preprint \cite{Seiberg:2025bqy} appeared on arXiv, exploring similar themes from a different perspective. We have revised the manuscript to better highlight the differences, especially in the examples considered.

\section{The 2D compact BF model and its global symmetries}\label{CBFT}

The topological gauge theory focused in this paper will mainly be the compact BF theory. So, in this section, we will review the compact BF model with some new points. The two-dimensional compact BF model is closely related to several other quantum field theories, including the $G/G$ gauged WZW model \cite{Blau:1993tv}, the 2D Schwinger model \cite{Witten:1978bc}, and the gauged linear sigma model (GLSM) \cite{Witten:1993yc}, among others. In fact, the compact BF theory can be viewed as a low-energy limit or a simplified sector of these theories.

More specifically, the 2D Schwinger model, after bosonization and neglecting the gauge field’s kinetic term, reduces to a $G/G$ gauged WZW model. 
Similarly, the low-energy bosonic sector of a GLSM with a Fano target can also be described by a $G/G$ gauged WZW model \cite{Witten:1993xi}. The connection between the $G/G$ model and the compact BF theory can be made explicit via a path-integral analysis.

Consider the abelian $G/G$ gauged WZW model, whose Lagrangian takes the form
\begin{equation}\label{BFL}
\frac{1}{8\pi} (\partial\varphi)^2 + \frac{N}{2\pi} A \wedge d\varphi,
\end{equation}
where $\varphi \sim \varphi + 2\pi$ -- this periodicity is what makes the theory “compact.”

Now perform the variable shift
\begin{equation}\label{PCV}
A \rightarrow A + \frac{1}{4N} \star d\varphi,
\end{equation}
where $\star$ stands for the Hodge dual, which is not a gauge transformation, but the Jacobian of the path-integral measure remains trivial. Under this change, the kinetic term for $\varphi$ is canceled, and the Lagrangian simplifies to
\begin{equation}\label{CBFL}
\frac{N}{2\pi} A \wedge d\varphi,
\end{equation}
revealing its topological nature. Upon integrating by parts, we obtain the familiar compact abelian BF form:
\begin{equation}\label{CBFL2}
\frac{N}{2\pi} \varphi F,
\end{equation}
where $F = dA$.

The compact BF theory possesses an explicit $\mathbb{Z}_N$ axial symmetry, acting as
$$
\varphi \rightarrow \varphi + \frac{2\pi q}{N}, \quad q=0,\ldots,N-1.
$$
In addition, this theory features a 1-form $\mathbb{Z}_N$ symmetry, with Wilson lines serving as the charged operators:
\begin{equation}\label{WL}
W[n]=\exp\left(in\int A\right).
\end{equation}

The generator for this 1-form symmetry is $e^{i\varphi}$,\footnote{Another way to see this is to put the theory on spacetime $\mathcal{M}= M \times  \mathbb{R}$ with compact space $M$, then it reduces to a quantum mechanical system with two variables $\varphi$ and $\tilde{\varphi}$\cite[Section 3]{Kapustin:2014gua}. More specifically, we write out the action to be 
\begin{equation}\nonumber
\int dtdx\frac{N}{2\pi}\varphi(\partial_0 A_1-\partial_1 A_0).
\end{equation}
EOM of $A_0$ gives $\partial_1\varphi=0$. Its solution is $\varphi:=\varphi(t)$. Then, plug this solution into the action, which becomes
\begin{equation}\nonumber
\frac{N}{2\pi}\int dt \ \varphi \frac{d\tilde{\varphi}}{dt},
\end{equation}
where $\tilde{\varphi}=\int_M A_1$.
}, and its effect on the  vacuum structure of the compact BF theory is the following. We read off from the partition function,
\begin{equation}\label{PFF}
   Z(N)=\int \mathcal{D}\varphi \mathcal{D}F \exp\left(i\frac{N}{2\pi}\int \varphi F \right),
\end{equation}
that has been derived in \cite[Section 7.1]{Blau:1993tv}.  To evaluate this expression, we begin with the EOMs derived from the Lagrangian:
\begin{equation}\label{EOM}
d\varphi=F=0,
\end{equation}
whose solutions are constant $\varphi_c$ and a flat gauge field $A_c$. 

On non-compact spacetime, we localize the theory to configurations with vanishing topological flux $\int F = 0$, expanding around classical backgrounds as
\begin{equation}\label{FL}
A = A_c + \delta A, \quad \varphi = \varphi_c + \delta \varphi,
\end{equation}
with fluctuations vanishing at spatial infinity. The axial symmetry suggests that $\varphi_c$ should be $\mathbb{Z}_N$-valued. The action would then be reduced to 
\begin{equation}\label{ReAc}
 S=\frac{N}{2\pi}\int  d(\varphi_c \delta A)+\delta \varphi \cdot \delta (dA),
\end{equation}
where the first term above is a total derivative and can be safely discarded. The partition function becomes
 \begin{equation}\label{PFFS}
   Z(N)=\int \mathcal{D}(\delta\varphi) \int \mathcal{D}(\delta F) \exp\left(i\frac{N}{2\pi}\int \delta\varphi \delta F \right)=\int^{2\pi}_0 d\varphi \ \delta\left(\frac{N}{2\pi} \left(\varphi-\varphi_c\right)\right).
\end{equation}
From this, we deduce a single vacuum configuration:
\begin{equation}
e^{i\varphi} = e^{i\varphi_c}, \quad F = 0,
\end{equation}
which matches the EOMs of the Lagrangian reformulated in terms of $F$ (instead of $A$). 

However, the presence of the $\mathbb{Z}_N$ axial symmetry suggests that there should be $N$ degenerate vacua. To make this structure manifest, we consider the theory on a compact spacetime $\Sigma$ (e.g., a Riemann surface), allowing for topologically nontrivial fluxes:
\begin{equation}\label{FLU}
\int_{\Sigma} F \in 2\pi \mathbb{Z}.
\end{equation}
Such fluxes contribute non-trivially to the path integral. Using the identity
  \begin{equation}\label{DPDF}
   \sum^{\infty}_{m=-\infty}\delta\left(\int_{\Sigma} F- 2\pi m\right)=\sum^{\infty}_{n=-\infty}\exp\left(in\int_{\Sigma} F\right) ,
\end{equation}
  we see that the effective action is modified in each topological sector labeled by $n$:
\begin{equation}\label{CBFL3}
   \frac{N}{2\pi}\varphi F+n F.
\end{equation}
The updated EOMs are
\begin{equation}\label{EOM2}
e^{i\varphi} = e^{i\left(\varphi_c-\frac{2\pi n}{N}\right)}, \quad F = 0,
\end{equation}
while the EOMs expressed in terms of $A$ remain unchanged. 

Finally, the partition function becomes a sum over sectors:
\begin{equation}\label{PFFS2}
Z(N) = \sum_{n \in \mathbb{Z}} \int \mathcal{D}\varphi \mathcal{D}F \exp\left(i \frac{N}{2\pi} \int \varphi F + i n \int F\right)
= \int_0^{2\pi} d\varphi \sum_{n \in \mathbb{Z}} \delta\left(\frac{N}{2\pi} \varphi - n\right).
\end{equation}
This gives rise to the vacuum condition
\begin{equation}\label{VSS}
e^{iN\varphi} = 1,
\end{equation}
which indeed admits $N$ distinct vacua, in agreement with the expected $\mathbb{Z}_N$ axial symmetry\footnote{Our counting of vacua implicitly assumes the spacetime manifold $T^2$, where the partition function directly encodes the number of vacua. See \cite{Fliss:2023dze,Fliss:2023uiv} for related discussion of $BF$ quantization.}.

The $l$-th vacuum sector can also be understood from the perspective of the theory defined on non-compact spacetime, by turning on a Wilson line at spatial infinity:
\begin{equation}\label{MWL}
W[l] = W[1]^l = \exp\left(il\int_L A\right),
\end{equation}
where $L$ denotes the worldline of the soliton. This Wilson line effectively shifts the vacuum from $e^{i\varphi} = e^{i\varphi_c}$ to $e^{i\varphi} = e^{i\left(\varphi_c-\frac{2\pi n}{N}\right)}$.

This shift can be made explicit by evaluating the path integral in non-compact spacetime, where the initial state is prepared by acting $W[l]$ on the ground state $\ket {\varphi_c=0}$:
\begin{align} \nonumber
       \bra{W[l]}\ket {W[l]} &=
\int D\varphi DF \exp\left(i\frac{N}{2\pi}\int \varphi F+i l\int  F\right) \nonumber \\
&=\int D\varphi DF \exp\left(i\frac{N}{2\pi}\int \left(\varphi+\frac{2\pi l}{N} \right) F\right) \nonumber \\
&= \int^{2\pi}_0 d\varphi \ \delta \left(\frac{N}{2\pi}\varphi+l\right)  ,\nonumber
\end{align}
Here we have used the relation $\oint_{\partial \Sigma} A = \int_{\Sigma} F$, which can be understood physically as arising from a soliton–anti-soliton pair separated by an infinite distance.

This calculation shows that the Wilson line acts as a generator of the axial 0-form $\mathbb{Z}_N$ symmetry. Each $W[l]$ creates a domain wall between distinct vacua labeled by the axial angle $\varphi$. Moreover, since the Wilson line is an extended operator, it can carry charge under the electric one-form $\mathbb{Z}_N$ symmetry.

In fact, there exists a mixed ’t Hooft anomaly between the axial 0-form symmetry and the electric 1-form symmetry. To see this, consider the vacuum state $\ket{p} = W[-p] \cdot \ket{0}$, where $\ket{0} \equiv \mathbf{1}$. Let $U_j = e^{i j\varphi}$ denote the topological operator measuring the one-form $\mathbb{Z}_N$ charge, with $j$ labeling elements of $\mathbb{Z}_N$. For instance,
\begin{equation}\nonumber
U_j W[l] U_j^{-1} = e^{-i\frac{2\pi j l}{N}} W[l],
\end{equation}
so $W[-l]$ carries one-form charge $l$ mod $N$. Acting on the vacuum, we find
\begin{equation}\label{CR}
U_j \cdot W[l] \ket{p} = e^{\frac{2\pi i j (p - l)}{N}} \ket{p - l}, \quad\quad W[l] \cdot U_j \ket{p} = e^{\frac{2\pi i j p}{N}} \ket{p - l}.
\end{equation}
Since $U_j$ and $W[l]$ generate the 1-form and 0-form symmetries, respectively, the non-commutativity in (\ref{CR}) reflects a mixed anomaly between them. This anomaly can also be derived via canonical quantization.

The action of 0-form symmetries by permutations on vacua (e.g.~it always factors through the $S_N$ subgroup of $U(N)$) and 1-form symmetries by phases is in fact a general property of semisimple TQFTs \cite{Gukov:2021swm}. The symmetry in this theory can also be understood from gauging of a $(-1)$-form magnetic $U(1)$ symmetry in the Maxwell theory. The pure $U(1)$ gauge has a $U(1)$ $(-1)$-form symmetry\footnote{This has been recently explored in, e.g., \cite{Aloni:2024jpb, Yu:2024jtk}. Earlier discussions of $(-1)$-form symmetries can be found in \cite{Cordova:2019jnf, Cordova:2019uob} and references therein.} and 1-form symmetry with mixed anomaly. One can gauge the $N$-fold cover of the $U(1)^{(-1)}$ via
\begin{equation}
    \mathbb{Z}_N \rightarrow U(1) \rightarrow U(1)^{(-1)}.
\end{equation}
This amount to adding
\begin{equation}\label{MOS}
\exp\left(iN\int \frac{\alpha}{2\pi} F\right),
\end{equation}
in the action and promoting $\alpha$ to a dynamical field.
After gauging, the trivially acting $\mathbb{Z}_N$ center leads to both a $\mathbb{Z}_N$ 0-form and 1-form symmetry. To be more precise, the two resulting symmetries are produced by two different mechanisms: On one hand, the topological lines, generating $\mathbb{Z}_N$ 0-form symmetry, come from the ``twisted sector'' of the spacetime-filling operators generating the $\mathbb{Z}_N$ center of the $(-1)$-form symmetry, and become genuine operators after gauging. On the other hand, the $\mathbb{Z}_N$ 1-form symmetry is generated as the quantum symmetry from gauging the $(-1)$-form symmetry\footnote{This quantum symmetry structure under gauging can be easily understood from a 3D SymTFT $\frac{N}{2\pi} \int B_2\wedge d\alpha$. The $(-1)$-form symmetry is realized by picking Dirichlet boundary condition for $\alpha$. Gauging the $\mathbb{Z}_N$ $(-1)$-form symmetry amounts to pick the Dirichlet boundary for $B_2$ instead, which becomes the background field for the quantum $\mathbb{Z}_N$ 1-form symmetry.}. The original $U(1)^{(1)}$ 1-form symmetry is broken due to the mixed anomaly with $U(1)^{(-1)}$. This also generalizes to higher dimensions, where gauging the N-fold cover of the magnetic $(d-3)$-form $U(1)$ symmetry of a $U(1)$ gauge theory leads to a BF theory with action $\frac{N}{2\pi}\int B_{d-2}\wedge F$.

 In the 2d case, alternatively, one can arrive at the same theory with the following ``dual" perspective. One first gauge the $(-1)$-form global symmetry, and then Higgs it with charge $N$ matter. This amount to adding (\ref{MOS})
in the action and promoting $\alpha$ to be a dynamical field. Then the theory will have a 0-form $\mathbb{Z}_N$ symmetry. Furthermore, since the $\alpha$ field interacts with the gauge field $A$ through the interaction in $(\ref{MOS})$, the 1-form $U(1)$ electric symmetry will be screened to a 1-form $\mathbb{Z}_N$ symmetry.

The compact BF theory can be embedded into a larger theory. In the next section, we introduce the GLSM and explore how it connects to the compact BF description. We will see that the GLSM retains (almost) the same axial 0-form symmetry, although the 1-form symmetry may be replaced by a different symmetry structure.

\section{Gauged linear sigma model for weighted projective space}\label{GLSM}

Gauged linear sigma models (GLSMs) were first introduced by Witten \cite{Witten:1993yc} to establish a correspondence between the nonlinear sigma model (NLSM) on a Calabi–Yau manifold and a two-dimensional orbifolded Landau–Ginzburg model. A few months later, Witten \cite{Witten:1993xi} employed GLSMs to demonstrate another remarkable correspondence: the quantum cohomology of the Grassmannian $Gr(k,N)$ is isomorphic to the Verlinde algebra of $U(k)_{N-k}$, which generalized and proved the conjecture made in \cite{Gepner:1990gr, Vafa:1991uz, Intriligator:1991an}.

For simplicity, consider a ${\mathcal{N}}=(2,2)$ $U(1)$ gauge theory with $n$ chiral matter superfields $\Phi_i$, each carrying a positive electric charge $w_i$ ($i=0,\dots,n-1$), such that $N = \sum_{i=0}^{n-1} w_i$. Each $\Phi_i$ takes the standard form $\Phi_i = \phi_i + \sqrt{2}\theta^\pm \psi_{i,\pm} + \ldots$, and the field strength superfield is $\Sigma = \sigma + \ldots$. The complexified Kähler parameter is $t = r - i\theta$, where $r$ is the Fayet–Iliopoulos (FI) parameter and $\theta$ is the theta angle.

The vacuum structure of the theory depends on the energy scale $\mu$, where the FI parameter runs logarithmically:
\begin{equation*}
    r = N \log \frac{\mu}{\Lambda},
\end{equation*}
with $\Lambda = \mu e^{-r(\mu)/N}$ the dynamical scale, and its complexified version given by ${\overline{\Lambda}} = \Lambda\, e^{i\theta/N}$.

Depending on the scale $\mu$, the theory has two distinct effective theories:
\begin{itemize}
    \item For $\Lambda \ll \mu \ll e\sqrt{r}$, the theory is weakly coupled, and flows to an NLSM with target space $\mathbb{CWP}(w_0,\dots,w_{n-1})$, a weighted projective space.
    \item For $\mu \ll \Lambda$, the theory reduces to an effective theory for $\Sigma$ with a twisted superpotential
    \begin{equation*}
        \int d^2\tilde{\theta}\, \frac{1}{2} \left[ 
        \left( -t + \sum_{i=0}^{n-1} w_i \log w_i \right) \Sigma - N\Sigma \left( \log \left( \frac{\Sigma}{\mu} \right) - 1 \right)
        \right] + \text{c.c.}
    \end{equation*}
    where $d^2\tilde{\theta}$ denotes the integral over the twisted superspace. The resulting vacuum equation is:
    \begin{equation*}
        \left( \prod_{i=0}^{n-1} w_i^{w_i} \right) \sigma^N = \overline{\Lambda}^N,
    \end{equation*}
    which determines the critical locus of the twisted superpotential. Integrating out the massive gauginos yields a bosonic effective theory, which matches the compact BF theory described in Section~\ref{CBFT}, upon identifying
    \[
        e^{i\varphi} := \sigma \left( \overline{\Lambda}^{-1} \prod_{i=0}^{n-1} w_i^{w_i} \right)^{1/N}.
    \]
\end{itemize}

Previous literature \cite{Morrison:1994fr, Pantev:2005zs} did not systematically analyze the symmetries of weighted projective spaces. For completeness, we now list the relevant symmetries.

\begin{itemize}
\item \textbf{In the NLSM} on target $\mathbb{CWP}(w_0,\ldots,w_{n-1})$, the theory exhibits:
    \begin{itemize}
        \item An R-symmetry:
        \begin{equation*}
            \mathbb{Z}_{2N} \times U(1)_V,
        \end{equation*}
        where $\mathbb{Z}_{2N}$ is the axial R-symmetry, and $U(1)_V$ is the vector R-symmetry.
        
        \item A flavor symmetry:
        \begin{equation*}
            G_F \subseteq SU(n),
        \end{equation*}
        which depends on the matter charges. For example, $\mathbb{P}^{N-1}$ has $SU(N)$ flavor symmetry. More generally, the unbroken flavor symmetry is the subgroup of $SU(n)$ that preserves the orbifold fixed points in $\mathbb{CWP}(w_0/g, \ldots, w_{n-1}/g)$, where $g = \mathrm{gcd}(w_0,\dots,w_{n-1})$.
        
        \item A discrete symmetry generated by solitons/domain walls \cite{Witten:1978bc}, associated to the matter fields. These generate at least a subgroup of the center $\mathrm{Z}(G_F)$ of the flavor symmetry group:
        \begin{equation*}
            \mathbb{Z}_{N/g}.
        \end{equation*}

        \item A possible 1-form symmetry:
        \begin{equation*}
            \mathbb{Z}_g.
        \end{equation*}
        Thus, the NLSM on projective space $\mathbb{P}^{N-1}$ (where $g=1$) does not have a 1-form symmetry, while a sigma model on $B\mathbb{Z}_N$ has a $\mathbb{Z}_N$ 1-form symmetry.
    \end{itemize}

\item \textbf{In the effective twisted Landau-Ginzburg theory} ($\mu \ll \Lambda$), the symmetries were analyzed in Section~\ref{CBFT} and include:
\begin{equation*}
    \mathbb{Z}_{2N} \times \mathbb{Z}_N,
\end{equation*}
where $\mathbb{Z}_N$ is the 1-form symmetry and $\mathbb{Z}_{2N}$ is the axial R-symmetry. A subtlety arises: the factor of $\mathbb{Z}_2$ in $\mathbb{Z}_{2N}$ acts trivially on the $\sigma$-vacua, since the vacuum must preserve Lorentz symmetry. As argued in the modern literature, e.g. \cite{Komargodski:2020mxz}, this $\mathbb{Z}_2$ should be regarded as a gauge symmetry. Hence, the physical axial R-symmetry is effectively $\mathbb{Z}_N$.
\end{itemize}

\noindent\textbf{Mixed 't Hooft anomaly.} In Section \ref{sec: general discussion}, we will argue that some non-genuine operators in the UV theory must be responsible for the anomaly observed in the IR. In this section, we test this general expectation through concrete examples. Recall from Section \ref{CBFT} that the compact BF theory exhibits a mixed 't Hooft anomaly between two $\mathbb{Z}_{N}$ symmetries. We now examine how this anomaly manifests in NLSMs.

We begin with two extreme cases. The first example is the projective space $\mathbb{P}^{N-1}$. One can verify that operators charged under the flavor symmetry are also charged under the 0-form $\mathbb{Z}_{N}$ axial R-symmetry. To see this, note that the $SU(N)$-charged operators take the form
\begin{equation} \label{GIO}
   \Phi W[1] := \phi(x_0) e^{i\int^{\infty}_{x_0} A_1 dx} + \ldots,
\end{equation}
where Witten showed in \cite[Section IV]{Witten:1978bc} that such operators can create a domain wall interpolating between neighboring vacua. As in Section \ref{CBFT}, one can similarly show that the operators in (\ref{GIO}) do not commute with the axial symmetry generators, $e^{ij\sigma}$. However, we must keep in mind that in the UV theory it is not possible to localize to a particular vacuum, so a fully precise derivation should take this into account.

Moreover, Witten demonstrated that the domain wall is a BPS object with mass proportional to the dynamical scale $\Lambda$. In the deep infrared, where $\mu \ll \Lambda$, both the matter field $\Phi$ and the gaugino are integrated out, and the operator at $x_0 = -\infty$ may be treated as trivial. In this regime, the expression (\ref{GIO}) effectively becomes a genuine line operator:
\begin{equation} \label{OWL}
   e^{i\int^{\infty}_{-\infty} A_1 dx}.
\end{equation}
This implies that the 0-form $SU(N)$ flavor symmetry in the UV is mapped to a 1-form $\mathbb{Z}_{N}$ symmetry in the IR, consistent with the symmetry structure of the compact BF theory.

The second example is the Abelian Higgs model with a charge-$N$ matter. Its IR limit is described by a 2D $\mathbb{Z}_{N}$ gauge theory, which is known to be equivalent to the topological coset model $U(1)_N / U(1)_N$. Thus, it is natural that they exhibit the same symmetry structure.

For a general weighted projective space, the analogs of the above operators take the form
\begin{equation}\label{WPWO}
    \Phi_i W[w_i], \quad\mathrm{for}\quad i = 0,\ldots,n-1,
\end{equation}
which are charged under a $\mathbb{Z}_{N/g}$ symmetry. Meanwhile, the charged objects under the remaining $\mathbb{Z}_{g}$ symmetry are Wilson lines of the form
\begin{equation}\label{WWL}
   W[l], \quad\mathrm{for}\quad l = 0,\ldots, g-1.
\end{equation}
As in the two previous examples, one finds that the operators in (\ref{WPWO}) also carry charges under axial symmetry. In the deep infrared, these operators become genuine Wilson lines. Hence, a 1-form $\mathbb{Z}_{N}$ symmetry emerges in the IR, matching the symmetry of the compact BF theory.

In the next section, we will explore the correspondence between the A-twisted NLSM on a weighted projective space and the $U(1)_N / U(1)_N$ theory. We will then propose a general correspondence between the quantum cohomologies of different target spaces that share the same axial symmetry.
   
\section{A correspondence between quantum cohomologies}\label{CBQH}

In \cite{Witten:1993xi}, it was shown that the quantum cohomology of $\mathbb{P}^{N-1}$ is isomorphic to the Verlinde algebra of $U(1)_N$. The quantum cohomology can be computed from the A-twisted NLSM on $\mathbb{P}^{N-1}$ \cite{Vafa:1991uz, Witten:1988xj} with $\overline{\Lambda} = 1$. In general, TQFTs are characterized by their structure constants and pairings. Hence, the isomorphism here refers to the fact that the two theories share identical ring structure constants, and their pairings match under an appropriate dictionary.

A first indication of this correspondence comes from the chiral ring relations in both theories:
\begin{equation}\label{VEQ}
\sigma^N = 1.
\end{equation}
This ensures the ring structures are the same. However, the pairings—defined via two-point correlators—differ between the two theories. Witten demonstrated that they are nevertheless related, by a path-integral argument involving the axial anomaly.

Specifically, the pairing in the Verlinde algebra can be recovered from that of the quantum cohomology via:
\begin{equation}\label{MCBF}
\left\langle{\mathcal{O}}_j(\sigma),{\mathcal{O}}_k(\sigma) \right\rangle_{U(1)_N} = \left\langle \sigma^{(N-1)(1-h)}, {\mathcal{O}}_j(\sigma), {\mathcal{O}}_k(\sigma)\right\rangle_{\mathbb{P}^{N-1}},
\end{equation}
where $h$ is the genus of the Riemann surface, and ${\mathcal{O}}_i(\sigma)$ are monomials in $\sigma$ corresponding to $U(1)$ representations. The extra insertion $\sigma^{(N-1)(1-h)}$ originates from the axial anomaly and was derived via path-integral methods in \cite{Witten:1993xi}.\footnote{For the quantum cohomology of $Gr(k,N)$, the corresponding insertion becomes $(\det \sigma)^{(N-1)(1-h)}$.}

Conversely, the pairing in quantum cohomology can be expressed in terms of the compact BF theory correlators:
\begin{equation}\label{PQH}
\left\langle {\mathcal{O}}_{j^\prime}(\sigma),{\mathcal{O}}_{k^\prime}(\sigma)\right\rangle_{\mathbb{P}^{N-1}} = \left\langle \sigma^{1-h},{\mathcal{O}}_{j^\prime}(\sigma),{\mathcal{O}}_{k^\prime}(\sigma) \right\rangle_{U(1)_N}.
\end{equation}

We now extend this correspondence to weighted projective spaces. Following the arguments in Sections~\ref{GLSM} and \ref{sec: general discussion}, we introduce a background gauge field $A^R$ for the axial symmetry. Due to the mixed ’t Hooft anomaly, the partition function transforms under axial shifts as:
\begin{equation}\label{MTGT}
Z(A^R + d\Lambda^R) = \exp\left(-i \int \Lambda^R \wedge dj^A\right) Z(A^R),
\end{equation}
with the axial current $j^A = \star d\varphi$. After the topological twist, only the ground states contribute, and the axial symmetry becomes anomalous, i.e., $dj^A \neq 0$. Defining the net axial charge violation across spacetime as $\Delta Q = \int dj^A$, we find:
\begin{equation}\label{AAC}
\Delta Q = 2(n-1)(1-h) + 2\left( \sum_{i=0}^{n-1} w_i \right) m,
\end{equation}
where the first term results from the mode counting of the A-twisted NLSM on a genus-$h$ surface: each matter fermion $\psi_-$ or $\bar{\psi}_+$ (spin 0) contributes one mode, and each $\psi_+$ or $\bar{\psi}_-$ (spin 1) contributes $h$ zero modes.\footnote{Alternatively, in the GLSM, the gauge field contributes $-2(1-h)$, and matter fields contribute $2n(1-h)$, consistent with the above.} Using axial charges listed in \cite[Chapter 16.2]{Hori:2003ic}, we reproduce this result. The second term in (\ref{AAC}) comes from the mixed anomaly discussed in Sections \ref{CBFT} and \ref{GLSM}. A more abstract and comprehensive discussion of this selection rule will be presented in Section \ref{sec: general discussion}.

To preserve most flavor symmetry but in a non-trivial background of the center symmetry, the allowed BPS operator is
\begin{equation}\label{GLB}
{\cal B}^k \coloneqq \prod_{i=1}^{n-1} \Phi_i^{k/N} W[k].
\end{equation}
This operator creates the background necessary to calculate the partition function in (\ref{MTGT}). The corresponding state is given by $\vert \Omega \rangle = \sum_{k \in \mathbb{Z}} \vert {\cal B}^k \rangle$.

Since $\sigma$ carries axial charge 2, its insertions can compensate the anomaly and lead to nonzero correlators:
\begin{equation}\label{MCBF2}
\left\langle \sigma^{(n-1)(1-h) + kN} \cdots \right\rangle_{\mathbb{CWP}(w_0, \ldots, w_{n-1})}.
\end{equation}
Writing $\sigma = |\sigma| e^{i\varphi}$, the axial rotation $\varphi \mapsto \varphi + \frac{2\pi}{N}$ implies
\begin{equation}
    e^{iN\varphi} = 1,
\end{equation}
which is the same vacuum constraint as in the compact BF theory. Furthermore, by tuning $\overline{\Lambda}^N = \prod_{i=0}^{n-1} w_i^{w_i}$, we can set $|\sigma| = 1$.

This leads us to a generalized correspondence:
\begin{equation}\label{MCBF2_alt}
\left\langle{\mathcal{O}}_j(\sigma),{\mathcal{O}}_k(\sigma) \right\rangle_{U(1)_N}=\left\langle \left(\prod^{n-1}_{l=0} w_l^{1-h}\right) \sigma^{(n-1)(1-h)}, {\mathcal{O}}_j(\sigma),{\mathcal{O}}_k(\sigma)\right\rangle_{\mathbb{CWP}(w_0,\ldots,w_{n-1})},   
\end{equation}
and the inverse relation:
\begin{equation}\label{PQH2}
\left\langle {\mathcal{O}}_{j^{\prime}}(\sigma),{\mathcal{O}}_{k^{\prime}}(\sigma)\right\rangle_{\mathbb{CWP}(w_0,\ldots,w_{n-1})}=\left\langle \left(\prod^{n-1}_{l=0} w_l^{-(1-h)}\right)\sigma^{(N-n+1)(1-h)},{\mathcal{O}}_{j^{\prime}}(\sigma),{\mathcal{O}}_{k^{\prime}}(\sigma) \right\rangle_{U(1)_N},  
\end{equation}
The overall factors $\prod_{l=0}^{n-1} w_l^{(h-1)}$ are determined from classical geometry \cite{Coates2006}. Importantly, up to an overall constant factor, this correspondence depends solely on $\sum_i w_i$, not on the individual values of $w_i$, suggesting that spaces with the same $\sum_i w_i$ share equivalent topological data, offering a novel classification of TQFTs. However, since flavor symmetries differ across these spaces, their equivariant deformations generally do not match. 

Of course, the ring relations of these target spaces, encoded by the twisted effective superpotential, can already be inferred from \cite{Witten:1993yc, Morrison:1994fr}. The quantum cohomology of gerby target spaces was studied in \cite{Pantev:2005zs}, while a mathematical treatment of the quantum cohomology of weighted projective spaces can be found in \cite{Coates2006}.

Our new observations are twofold. First, we show that the quantum cohomology of $\mathbb{CWP}(\vec{w})$ is also isomorphic to the Verlinde algebra of $U(1)_N$, where $N=\sum_i w_i$, thereby extending the well-known correspondence between the quantum cohomology of $\mathbb{P}^{N-1}$ and the Verlinde algebra of $U(1)_N$. Second, this Verlinde-algebra interpretation leads to new correspondences among the resulting quantum cohomology rings.

This correspondence can also be derived from the geometry of the instanton moduli space, as discussed in \cite{Morrison:1994fr}. We review and translate the mathematical formulation of the quantum cohomology of weighted projective spaces into our language in the Appendix~\ref{appendix:a}.

\vspace{1em}
\noindent \textbf{Example: $\sum_i w_i = 4$}. Possible target spaces include $\mathbb{P}^3$, $\mathbb{CWP}(1,1,2)$, $\mathbb{CWP}(1,3)$, $\mathbb{CWP}(2,2)$, and $B\mathbb{Z}_4$. Their ring relations are all $\sigma^4 = 1$ if we set $\overline{\Lambda}^4 = 1$, 4, 27, 16, and 256, respectively. The small quantum cohomology is fully specified by its ring structure and pairings, which, following Eq.~(\ref{PQH2}), can be computed from the correlators for the Verlinde algebra of $U(1)_4$. For instance, for $\mathbb{CWP}(2,2)$ on $S^2$:
\begin{equation}\nonumber
\left\langle \sigma^j\sigma^k\right\rangle_{\mathbb{CWP}(2,2)}=\frac{1}{4}\left\langle \sigma^3\sigma^j\sigma^k\right\rangle_{U(1)_4}.
\end{equation}
The right-hand side can be computed by inserting $e^{i(3+j+k)\varphi}$ into the path-integral of the compact BF model:
\begin{equation}\nonumber
\left\langle \sigma^3\sigma^j\sigma^k\right\rangle_{U(1)_4}=\int^{2\pi}_0 d\varphi \sum_{n\in \mathbb{Z}} e^{i(3+j+k)\varphi} \delta\left(\frac{4}{2\pi} \varphi-n\right).
\end{equation}
They are zero unless $j+k=1$ mod 4, which is precisely the selection rule (\ref{AAC}) for NLSM on $\mathbb{CWP}(2,2)$ with $h=0$.

\section{3D Chern-Simons-matter theory for weighted projective space}
\label{CSMT}

In this section, we discuss 3D Chern-Simons-matter theories. As shown in \cite{Blau:1993tv}, the compact BF theory is equivalent to the 3D Chern-Simons theory compactified on $\Sigma \times S^1$, where $\Sigma$ denotes a 2D spacetime. This equivalence becomes clearer from the symmetry perspective: the $\mathbb{Z}_N$ 1-form symmetry of the 3D CS theory splits into a $\mathbb{Z}_N$ 0-form and a $\mathbb{Z}_N$ 1-form symmetry on $\Sigma \times S^1$. Moreover, the mixed 't Hooft anomaly in the resulting 2D theory originates from the 3D 1-form symmetry anomaly \cite{Gaiotto:2014kfa}.

Analogous to the 2D case, one can embed the 3D abelian CS theory into a 3D Chern-Simons-matter theory targeting a weighted projective space \cite{Gu:2021yek}. In this construction, new input parameters appear as bare Chern-Simons levels: the gauge CS level is given by $k = \sum_{i=0}^{n-1} \frac{w_i^2}{2}$, and the gauge-R mixed CS level by $k^{gR} = -\frac{n}{2} + 1$. We turn off other CS levels, including those involving flavor symmetries. The phase structure of the theory depends on the 3D FI parameter $\zeta$:
\begin{itemize}
    \item For $0 \ll \zeta$, the vacuum describes a 3D NLSM with target $\mathbb{CWP}(w_0,\ldots,w_{n-1})$;
    \item For $\zeta \ll 0$, the theory flows to a 3D $U(1)_N$ pure Chern-Simons theory, with effective gauge CS level $N = \sum_{i=0}^{n-1} w_i^2$, and effective gauge-R mixed CS level $-n + 1$ \cite[eqs.~(2.6), (2.12)]{Intriligator:2013lca}.
\end{itemize}

The appearance of a topological phase in 3D reflects the presence of \textit{non-genuine operators} in the UV theory, which flow to topological line operators that are charged under themselves. By placing the 3D theory on $\Sigma \times S^1$, one obtains a Witten-type TQFT that connects to quantum K-theory \cite{Kapustin:2013hpk, Gukov:2015sna, Ruan2018, Zhang2022}. \footnote{To recover Givental-Lee’s quantum K-theory, the bare gauge-R CS level must be chosen as $k^{gR} = \frac{N}{2}$ \cite{Ueda:2019qhg, Gu:2024mqk}, though the underlying selection rule becomes less transparent in this case.}

The corresponding vacuum equation (or ground ring relation) reads \cite{Gu:2021yek, Aganagic:2001uw}:
\begin{equation}\label{QKVE}
    \prod_{i=0}^{n-1} (1 - x^{w_i})^{w_i} = e^{-\zeta},
\end{equation}
where $x$ denotes the Wilson loop around the spatial $S^1$. For generic $\zeta$, the vacuum equation depends explicitly on the weight data $w_i$, even when $N = \sum w_i^2$ is fixed. This means that the quantum K-theories of different weighted projective spaces with the same $N$ are not isomorphic, due to the different orbifold structures on $S^1$ \cite{Gu:2021yek}, though the topological twist only applies to $\Sigma$. However, in the $\zeta \to -\infty$ limit, the vacuum equation simplifies:
\begin{equation}\label{QKVE2}
    x^N = (-1)^{\sum w_i} e^{-\zeta},
\end{equation}
which can be rewritten as $\tilde{x}^N = 1$ by redefining $x$. This is expected as the theory flows to a pure CS phase. We thus conjecture:  
\begin{center}
    \textit{Quantum K-theories of different weighted projective spaces with the same $\sum w_i^2$ are isomorphic in the $\zeta \to -\infty$ limit.}
\end{center}

To close this section, we revisit the extra insertion introduced in Section~\ref{CBQH} from the perspective of the 3D theory at $\zeta \ll 0$, where the low-energy effective action becomes:
\begin{equation}\label{EFCS}
    S_{\mathrm{eff}} = \int  \left( \frac{N}{4\pi} A \wedge dA - \frac{n-1}{4\pi} A \wedge dA^R \right),
\end{equation}
with $A^R$ the background field for the $U(1)$ axial R-symmetry. Integrating out heavy matter fields shifts the CS levels accordingly. Upon compactifying on $\Sigma \times S^1$, Blau and Thompson \cite{Blau:1993tv} showed that the 3D CS theory reduces to the 2D compact BF model:
\begin{equation*}
    \int_{\Sigma} \frac{N}{2\pi} \varphi F.
\end{equation*}
Extending their method, (\ref{EFCS}) reduces to:
\begin{equation}\label{EFBF}
    \int_{\Sigma} \left( \frac{N}{2\pi} \varphi F - \frac{n-1}{4\pi} \varphi F^R \right).
\end{equation}
Although the vacuum equation for $\varphi$ remains unchanged, the axial symmetry is explicitly broken by the second term, in agreement with its role in NLSM. Moreover, topological fluxes, as in (\ref{FLU}), can be incorporated by modifying the functional measure: $\mathcal{D}F \rightarrow \sum_{m \in \mathbb{Z}} m \Omega$ with $\int_\Sigma \Omega = 2\pi$. The resulting correlators take the form:
\begin{equation}\label{CldCS}
    \left\langle e^{i\varphi j} \right\rangle = c \sum_{m \in \mathbb{Z}} \int_0^{2\pi} d\varphi \ 
    e^{i\varphi \left( j + (n-1)(g-1) - Nm \right)}.
\end{equation}
Hence, only the correlators satisfying the selection rule:
\begin{equation*}
    j = (n-1)(1 - g) + Nm
\end{equation*}
are non-vanishing, reproducing the result of NLSM.

Interestingly, the second term in (\ref{EFBF}),
\begin{equation*}
    -\int_\Sigma \frac{n-1}{4\pi} \varphi F^R,
\end{equation*}
can be interpreted as the insertion
\begin{equation*}
    e^{i\varphi (n-1)(g-1)}
\end{equation*}
in the path integral of the compact BF theory, since it localizes on the constant mode of $\varphi$. Thus, the correlators in (\ref{CldCS}) align with those in geometric theory. Furthermore, for the projective space $\mathbb{P}^{N-1}$, it was shown in \cite{Ueda:2019qhg} that the quantum K-theory with these CS levels coincides with quantum cohomology upon identifying $1 - x \mapsto \sigma$, even without taking a strict 2D limit. This result generalizes to the non-abelian case, originally proposed in \cite{Kapustin:2013hpk}. However, for general weighted projective spaces, one must take the 2D limit to recover quantum cohomology, and a phenomenon known as “decomposition” arises in this process \cite{Pantev:2005zs, Gu:2021yek}.

\section{Mixed 't Hooft anomalies and RG flow}
\label{sec: general discussion}
In this final section, we return to the central conceptual question posed in the introduction:

\begin{quote}
\emph{Given an IR TFT with a mixed 't Hooft anomaly -- especially one involving an emergent symmetry -- what constraints, if any, can be placed on the structure of its ultraviolet completion?}
\end{quote}

Somewhat unexpectedly, such an anomaly imposes concrete, structural requirements on the UV theory. Key observation among these is the necessity for \emph{non-genuine} operators—operators attaching to one-dimensional higher objects that are not charged under any manifest UV symmetry, yet play a pivotal role in realizing emergent IR symmetry generators and matching the associated anomaly.

This insight emerges from a remarkably consistent pattern observed across the concrete models analyzed in previous sections. To motivate our general discussion, let us briefly recall our representative example from two perspectives:

\begin{itemize}
    \item \textbf{2D Compact BF theory and GLSM origins:} The IR compact BF theory exhibits a mixed 't Hooft anomaly between an axial $\mathbb{Z}_N$ 0-form symmetry and an electric $\mathbb{Z}_N$ 1-form symmetry (Eq.~\eqref{CR}). This structure descends from UV theories such as gauged linear sigma models (GLSMs) flowing to nonlinear sigma models on $\mathbb{P}^{N-1}$ or on weighted projective spaces $\mathbb{C}\mathbb{WP}(\vec{w})$. In the UV, operators like $\Phi_i W[w_i]$ (Eq.~\eqref{WPWO})—charged under the flavor symmetry but accompanied by Wilson lines—are intrinsically non-genuine. Under RG flow, they become genuine Wilson line operators $W[l]$ that carry charge under the emergent 1-form symmetry.
    
    \item \textbf{3D Chern–Simons–Matter theory:} In the deep IR of a 3D $U(1)_N$ Chern–Simons theory (arising from a matter theory with $\zeta \ll 0$), one finds topological line operators that are charged under themselves. This self-charging behavior is only compatible with odd-dimensional spacetimes and implies the existence of an emergent symmetry with a 't Hooft anomaly. 
\end{itemize}

Taken together, this suggests a more general mechanism: the data of a mixed IR anomaly encode the imprint of UV operators that may not be associated with any manifest symmetry but whose presence is required for anomaly matching and the consistent emergence of IR symmetries.

This motivates the key question that guides this section:

\begin{quote}
\emph{Given an IR TQFT with a mixed 't Hooft anomaly between $p$-form global symmetry $G^{(p)}_{\text{IR}}$ and an emergent q-form global symmetry $G^{(q)}_{\text{IR}}$, what can be deduced about the UV operator content—particularly when the UV symmetry structure differs, i.e., when $G^{(r)}_{\text{UV}} \neq G^{(q)}_{\text{IR}}$?}
\end{quote}

\paragraph{Mixed anomalies and operators.} Let us start with a minimal review of the mixed 't Hooft anomaly. It can be characterized through its effect on the partition function. Consider a quantum field theory with a $p$-form global symmetry $G^{(p)}$. When we turn on the background field $B_{p+1}$ for this symmetry, the partition function $Z[B_{p+1}]$ is said to possess a 't Hooft anomaly if it fails to be invariant under background gauge transformations:
\begin{equation}
    Z[\tilde{B}_{p+1}] \neq Z[B_{p+1}]
    \label{eq:anomaly-partition}
\end{equation}
where $\tilde{B}_{p+1} = B_{p+1} + \delta\lambda_{p}$.

A mixed 't Hooft anomaly between two symmetries $G^{(p)}$ and $G^{(q)}$ arises when the partition function transforms nontrivially under gauge transformations of one symmetry in the presence of the other's background field:
\begin{equation}
    Z[\tilde{B}_{p+1}, B_{q+1}] \neq Z[B_{p+1}, B_{q+1}], \quad
    \text{while} \quad
    Z[\tilde{B}_{p+1}, B_{q+1}=0] = Z[B_{p+1}, B_{q+1}=0]
    \label{eq:mixed-anomaly-partition}
\end{equation}
This signifies that the symmetries cannot be simultaneously gauged. If the theory of interest is an IR theory, it can happen that the $q$-form global symmetry $G^{(q)}$ is not a fundamental symmetry but an emergent one. There can be a ${G^{(r)}}$ symmetry in the UV theory, which is different from $G^{(q)}$. For the ease of reading, we will label symmetries accordingly as $G^{(r)}_{\text{UV}}$ and $G^{(q)}_{\text{IR}}$.

In operator language, this mixed anomaly manifests as non-commutativity of topological symmetry generators. Denoting the topological operators for $G^{(p)}$ and $G^{(q)}$ as $U_{G^{(p)}}(M^{d-p-1})$ and $U_{G^{(q)}}(\tilde{M}^{d-q-1})$ respectively, the anomaly implies:
\begin{equation}
    [U_{G^{(p)}}(M^{d-p-1}), U_{G^{(q)}}(\tilde{M}^{d-q-1})] \neq 0
    \label{eq:anomaly-commutator}
\end{equation}
when the submanifolds $M$ and $\tilde{M}$ are appropriately linked. This non-commutativity signifies that the topological operators of one symmetry carry charge under the other, as explicitly demonstrated in the BF model in Section \ref{CBFT}.

When $G^{(q)}_{\text{IR}}$ is emergent in the IR TQFT, this algebraic structure necessitates specific UV precursors. As motivated by our examples, these precursors take the form of \emph{non-genuine operators}.

\subsection*{Two mechanisms for emergent symmetries}
We now systematize the insights from our examples into two universal mechanisms that connect UV operator content to emergent IR anomalies. Both mechanisms crucially involve non-genuine operators.

A general belief is that RG flow builds a homomorphism between the UV and IR symmetry group. However, this needs to be treated carefully by looking at both topological operators as well as charged objects. Symmetry generators are topological operators and thus are protected under the RG-flow \cite{Gaiotto:2014kfa}, while their charged operators, on the other hand, are not protected. However, it does not mean that the symmetry can not ``disappear." For example, a well-known situation is spontaneous symmetry breaking via giving vacuum expectation value (vev) for some scalars. Another possible scenario that we will mainly focus on in this paper, although it is less appreciated before, is that all non-trivial charged operators of the group $G$ flow to the trivial representations, i.e. fully non-faithful in the IR (such as they all become very heavy), then the IR theory will not detect the $G$ symmetry at all. The action of G becomes trivial in the IR. Formally, write it as
\begin{equation}\label{RGF}
   {\cal RG}:  G_\text{UV}  \leadsto 1_\text{IR}, 
\end{equation}
where ${\cal RG}$ denotes the operation of RG-flow. More generally, we expect ${\cal RG} : G_\text{UV}\leadsto (G_0)_\text{IR}$, where $(G_0)_\text{IR}$ is a quotient of $G_\text{UV}$.\footnote{When there is spontaneous symmetry breaking, we can still discuss a $G$-symmetric IR phase by including all the vacua.}   

A more interesting situation is that a different form symmetry, $G^{(q)}$, emerges in the IR TQFT,\footnote{A more general situation would be a extension of $(G_0)_\text{IR}$, a quotient of $G_\text{UV}$, by the emergent symmetry $G^{(q)}_\text{IR}$, possibly to a higher group. 
 We will not deal with this more general scenario but instead assume that $G_0$ is trivial.} namely
\begin{equation}\label{RGF3}
   {\cal RG}: {G^{(r)}_\text{UV}}\leadsto G^{(q)}_\text{IR}.
\end{equation} 
Since we are interested in how to use $G^{(q)}_\text{IR}$ to probe the UV theory, thus while it is possible that the generators for $G^{(q)}_\text{IR}$ are truly emergent (i.e.~not descending from operators in UV), we will focus on the scenario that they actually originate from the UV and are related to the $G^{(r)}_\text{UV}$ symmetry.

\paragraph{Mechanism I: UV non-genuine operators $\leadsto$ IR symmetry-charged operators.}
The first mechanism involves that the operator charged under $G^{(r)}_\text{UV}$ are related to the ${G^{(q)}_\text{IR}}$-charged operator
\begin{equation}\label{RGFO1}
   {\cal RG}: \mathbb{ O}_{G^{(r)}_\text{UV}} \leadsto {\cal O}_{G^{(q)}_\text{IR}} .
\end{equation}
Here, we use a special notation for UV charged operators so that $\mathbb{ O}_{G^{(r)}_\text{UV}}$ includes not only genuine charged operators but also non-genuine operators. 
As we focus on the case that $G^{(r)}_\text{UV}$ is fully absent in the IR, where all the genuine operators in $\mathbb{O}_{G^{(r)}_\text{UV}}$ decoupled, the IR charged operator ${\cal O}_{G^{(q)}_\text{IR}}$ can only inherit from a non-genuine operator. This non-genuine operator can be denoted as ${\cal O}_{G^{(r)}_\text{UV}}$, attaching to another operator
\begin{equation}\label{NGO}
   {\cal O}_{G^{(r)}_\text{UV}}(\Gamma){\cal O}(\cal C),
\end{equation}
where $\partial {\cal C}=\Gamma$ (or $\partial \Gamma= \cal C$), so $\mathrm{dim} {\cal C}=\mathrm{dim}\Gamma+1$ (or $\mathrm{dim} {\cal C}=\mathrm{dim}\Gamma-1$). Therefore, the RG-flow for charged operators in (\ref{RGFO1}) should be replaced by a more precise form 
\begin{equation}\label{RGFO2}
   {\cal RG} : {\cal O}_{G^{(r)}_\text{UV}}\leadsto 1, \quad {\cal RG}:\left({\cal O}_{G^{(r)}_\text{UV}}(\Gamma){\cal O}({\cal C})\right)\leadsto{\cal O}_{G^{(q)}_\text{IR}}({\cal C}).
\end{equation}

Having discussed charged operators, let us now turn to the UV counterpart of topological operators of symmetry $G^{(q)}_\text{IR}$. Again given $r\neq q$, these symmetry operators cannot directly inherit from the topological operators of $G^{(r)}_\text{UV}$. Assume we have  another symmetry $G^{(p)}$ with the charged operators denoted as ${\cal O}_{G^{(p)}}$ which has $\mathrm{dim} {\cal O}_{G^{(p)}}\ge p$ for being charged. It can flow to the topological operators of $G^{(q)}_\text{IR}$, namely
\begin{equation}\label{RGFOG}
   {\cal RG} : {\cal O}_{G^{(p)}}\leadsto U_{G^{(q)}_\text{IR}}(M^{d-q-1}),
\end{equation}
 which is possible if $\mathrm{dim}{\cal O}_{G^{(p)}}+q+1=d$. It of course can happen that there is a non-trivial kernel of the map of ${\cal RG}$, namely there are UV operators map to the trivial one in the IR. The emergent topological operators indicate that there is a mixed anomaly in the IR since they are charged under each other. This is not rare; for example, the compact BF theory has a mixed 't Hooft anomaly. See Section \ref{CBFT} for the details of the 2D case. However, we already have the $G^{(q)}$-charged operators ${\cal O}_{G^{(q)}}$ from the UV, then it suggests the redundancy of topological operators of $G^{(p)}$ symmetry, namely:
\begin{equation}\label{RGFOG2}
    \{U_{G^{(p)}}(M^{d-p-1})\} \subseteq   \{  {\cal O}_{G^{(q)}}\} ,
\end{equation}
which can happen if $\mathrm{dim} {\cal C}+p+1=d$.

\paragraph{Mechanism II: UV symmetry-charged operators $\leadsto$ IR symmetry generators.}
The second choice is that the $G^{(r)}_\text{UV}$ charged operators are mapped under the RG-flow to the topological symmetry operators of ${G^{(q)}_\text{IR}}$ in the IR,
\begin{equation}\label{RGFO3}
   {\cal RG} : {\cal O}_{G^{(r)}_\text{UV}}\leadsto U_{G^{(q)}_\text{IR}}(M^{d-q-1}).
\end{equation}
To make this symmetry faithfully act on the theory, we require that it has non-trivial charged operators. In TQFT, it is very typical that topological operators are also charged operators. This imposes a condition to the dimension of the operators in $d$-dimensional QFT, that is
\begin{equation}\label{TDFDC}
   2(d-q-1)=d-1 \Rightarrow 2q+1=d.
\end{equation}
So $d$ can only be an odd number in this case. For example, in 3D Chern-Simons matter theory, the 1-form symmetry can appear in the topological phase, where its topological operators are also charged under themselves.
To make everything possible, some charged operators of ${\cal O}_{G^{(r)}}$ should be replaced by the non-genuine operator as in (\ref{NGO}) unless $r=q$. On the other hand, if $r=q$, the charged operators and the topological operators of the UV symmetry will organize themselves into the topological operators of the emergent IR symmetry and we do not absolutely need the non-genuine operators in this situation.

Another possibility is that the TQFT has a mixed 't Hooft anomaly with another symmetry $G^{(p)}$. This tells us that ${\cal O}_{G^{(r)}}$ must be charged under $G^{(p)}$ too, but how can this happen? Like in the first chocie as well, ${\cal O}_{G^{(r)}}$ should be replaced by a non-genuine operator 
\begin{equation}\label{NGO2}
   {\cal O}_{G^{(r)}}(\Gamma){\cal O}(\cal C),
\end{equation}
where $\partial {\cal C}=\Gamma$ (or $\partial \Gamma= \cal C$), so $\mathrm{dim} {\cal C}=\mathrm{dim}\Gamma+1$ (or $\mathrm{dim} {\cal C}=\mathrm{dim}\Gamma-1$). In order to make this possible, we should impose that
\begin{equation}\label{COC}
   r\le \mathrm{dim}\Gamma, \quad p\le \mathrm{dim}{\cal C}    
\end{equation}
since the $p$-form symmetry can only act on operators of dimensions no less than $p$. Furthermore, in order to become the topological operators of symmetry $G^{(q)}$, it requires $q+\mathrm{dim} {\cal C}+1= d$.

\subsection*{Witten-type TQFTs}

When a quantum field theory flows to a gapped phase, the infrared dynamics is usually captured by a topological field theory. Suppose again that the IR theory exhibits two symmetries, $G^{(p)}$ and $G^{(q)}_{\mathrm{IR}}$, with a mixed ’t Hooft anomaly between them. In such cases, the ground states can transform under $G^{(p)}$, while domain walls—frozen in the gapped limit—can carry charges under $G^{(q)}_{\mathrm{IR}}$. If the two symmetries coincide, the system exhibits a self-anomaly; otherwise, the anomaly encodes a nontrivial interplay between them.

However, not all TQFTs arise purely as IR phases. A different type of topological theory—Witten-type TQFT—can be constructed from a UV QFT via a topological twist, typically involving an vector/axial R-symmetry. Unlike Schwarz-type TQFTs, Witten-type TQFTs retain memory of massive or non-genuine operators in the original theory, even after supersymmetric twisting.

This distinction becomes crucial when tracking mixed ’t Hooft anomalies. After the twist, some massive non-BPS states (collectively denoted $\mathcal{H}_M$) are projected out; nevertheless, their remnant effects persist in the anomaly structure. In particular, non-genuine operators—such as certain baryonic combinations attached to Wilson lines—can encode the anomaly in the topologically twisted theory. These operators may be formally invisible in the IR, but remain essential for reproducing the anomaly under background symmetry transformations.

For instance, consider $G = G_0 = SU(N)$, with fundamental fields $\Phi_i$. A composite operator of the form
\begin{equation}\label{DCO}
 {\cal B}=(\det \Phi)^{\frac{1}{N}}W[1],
\end{equation}
is charged under the center of $SU(N)$ and can be used to build domain-wall-like insertions. Acting on a fixed vacuum $|0\rangle$, we may define a tower of states $|B^k\rangle = B^k |0\rangle$, and construct a $G$-invariant combination by summing over $k \in \mathbb{Z}$. This defines the state
\begin{equation}\label{FNSS}
    \vert \Omega \rangle=\sum_{k\in \mathbb{Z}}\vert {\cal B}^k \rangle,
\end{equation}
and the partition functions reads
\begin{equation}\label{UVPF}
    Z=\langle \Omega_f\vert  \Omega_i \rangle=\int  {\cal D \varphi} \  e^{-S(\varphi)}.
\end{equation}
The anomaly effect can be seen from that under a $G^{(0)}$ background gauge transformation, the insertion $|\Omega\rangle$ transforms nontrivially:
\begin{equation}\label{PIFN}
   \langle \Omega_f\vert G^{(p=0)}\cdot \vert \Omega_i \rangle=\int {\cal D}\varphi \  e^{-S\left(\varphi+\frac{2\pi}{N}\right)}=g\int {\cal D}\varphi \  e^{-S\left(\varphi\right)}=g\sum_{m\in \mathbb{Z}}\int {\cal D}\varphi \ g(e^{i\varphi}) e^{im N\varphi}f_m,
\end{equation}
where $N$ is a parameter that relates with the axial symmetry, $g$ is an anomaly phase factor that comes from $G^{(p=0)}\cdot\vert 0\rangle=g\vert 0\rangle$, $g(e^{i\varphi})$ is function of the physical variable $e^{i\varphi}$, and we have used the Fourier series in the last equality since only constant modes left in a TQFT. This structure leads to a selection rule: only those observables whose $G^{(0)}$ charge cancels the anomaly can have nonzero correlators.

From this point of view, the Witten-type TQFT serves as a bridge: it encodes the UV origin of the anomaly via topologically twisted non-genuine operators, while the IR topological gauge theory realizes the same anomaly through symmetry generators and domain wall sectors. Hence, the anomaly data defines an equivalence class of twisted theories, which all flow to the same IR TQFT, despite distinct UV constructions.

\section*{Acknowledgements}
We thank Andreas Karch, David Tong, and Yaoxiong Wen for useful discussions. The research of WG is funded by the National Natural Science Foundation of China (NSFC) with Grant No.12575077. The work of DP is partly supported by research grant 42125 from VILLUM FONDEN and Simons Collaboration on ``New Structures in Low-dimensional topology.'' XY is supported by NSF grant PHY-2310588.

\appendix

\section{Quantum cohomlogy of weighted projective spaces}\label{appendix:a}
In this section, we briefly outline the results of quantum cohomology of a weighted projective space $\mathbb{CWP}(w_0,\ldots,w_{n-1})$ studied in a math paper \cite{Coates2006}. 

The authors in \cite{Coates2006} introduced more generators and more relations than our discussion on $\sigma$-vacua, where it only has one generator, $\sigma$, and one ring relation, $\left(\prod^{n-1}_{i=0}w_i^{w_i}\right) \sigma^N=\overline{\Lambda}^N$. Their presentation is more or less for a NLSM. More specifically, their generators are
\begin{equation}\nonumber
    \textbf{1}_{f_1}(=\textbf{1}_{0}), \ \ldots ,\ \textbf{1}_{f_k}, \ \sigma,
\end{equation}
where $f_1(=0),\ldots,f_k$ be the elements of the set
\begin{equation}\nonumber
    F=\left\{\frac{k}{w_i}\mid 0 \leq k< w_i, 0\leq i\leq n\right\}
\end{equation}
arranged in increasing order, and set $f_{k+1}=1$.  The generator $\textbf{1}_{f_k}$ for $k\neq 1$ represents a twisted sector of orbifolds, which is also called Chen-Ruan cohomology in the mathematics literature. The relations are generated by
\begin{equation}\label{GWR}
\left(\prod_{b\in I_{f_i}} w_b \cdot\sigma \right)\textbf{1}_{f_j}=\overline{\Lambda}^{N(f_{j+1}-f_j)}\textbf{1}_{f_{j+1}},
\end{equation}
where $I_{f}\coloneqq \left\{i\mid w_i \cdot f\in \mathbb{Z} \right\}$. In particular, 
\begin{equation}\label{GWR2}
\left(\prod^{n-1}_{i=0} w_i^{w_i}\right)  \sigma ^N=\overline{\Lambda}^{N}.
\end{equation}
(\ref{GWR}) suggests that the twisted sectors can be expressed in terms of the functions of $\Lambda$ and $\sigma$. In other words, the small quantum orbifold cohomology is generated by $\sigma$, with a single relation (\ref{GWR2}).

We end this section by giving an explicit example: $\mathbb{CWP}(1,3)$.  In this case, $F=\left\{\frac{1}{3}, \frac{2}{3} \right\}$, and $f_1=0$, $f_2=\frac{1}{3}$, and $f_3=\frac{2}{3}$. The generators are:
\begin{equation}\nonumber
  \textbf{1}_{0},\ \textbf{1}_{\frac{1}{3}}, \ \textbf{1}_{\frac{1}{3}}, \ \sigma
\end{equation}
with relations:
\begin{equation}\nonumber 3\sigma^2\textbf{1}_{0}=\textbf{1}_{\frac{1}{3}}\overline{\Lambda}^{\frac{4}{3}}, \quad 3\sigma\textbf{1}_{\frac{1}{3}}=\textbf{1}_{\frac{2}{3}}\overline{\Lambda}^{\frac{4}{3}}, \quad 3\sigma \textbf{1}_{\frac{2}{3}}=\textbf{1}_{0}\overline{\Lambda}^{\frac{4}{3}}.
\end{equation}
From the above, it is easy to see that the small quantum cohomology of $\mathbb{CWP}(1,3)$ can be generated by $\sigma$ with only one relation: $3^3 \sigma^4=\overline{\Lambda}^4$.

\end{document}